\documentclass{hotnets17}
\pdfoutput=1
\usepackage{times}
\usepackage{fullpage}
\usepackage{hyperref}
\hypersetup{pdfstartview=FitH,
  pdfpagelayout=SinglePage,
  colorlinks=true,      
  linkcolor=blue,       
  citecolor=magenta,    
  filecolor=cyan,       
  urlcolor=blue          
}
\usepackage{graphicx, xspace, outlines,url,grffile,enumerate,listings,appendix,balance}
\usepackage[normalem]{ulem}
\usepackage{units}
\usepackage{xcolor}
\usepackage[footnotesize,it]{caption}
\usepackage{paralist}
\usepackage{amsmath}
\usepackage{amsfonts}
\usepackage{textcomp}
\usepackage{microtype}
\usepackage{siunitx}
\usepackage{cite}
\usepackage{mathptmx}
\usepackage{outlines}
\usepackage{enumitem}
\setitemize{itemsep=1pt,topsep=1pt,parsep=1pt,partopsep=1pt}
\newcommand{\bi}{\begin{itemize}}
\newcommand{\ei}{\end{itemize}}

\newcommand{\eg}{{\it e.g.,}\xspace}
\newcommand{\ie}{{\it i.e.,}\xspace}
\newcommand\eat[1]{}

\newcommand{\systemname}{ThrottleBot\xspace}
\newcommand{\bnr}{IMR\xspace}
\newcommand{\bnrs}{IMRs\xspace}
\newcommand{\mbnr}{MIMR\xspace}
\newcommand{\allnotes}[1]{}
\renewcommand{\allnotes}[1]{\textit{#1}}

\newcommand{\notemichael}[1]{\allnotes{\textcolor{purple}{[Michael: #1]}}}
\usepackage{array}
\usepackage{ragged2e}
\newcolumntype{P}[1]{>{\RaggedLeft\hspace{0pt}}p{#1}}
\usepackage{enumitem}
\setitemize{noitemsep,topsep=0pt,parsep=0pt,partopsep=0pt}
\DeclareSymbolFont{extraup}{U}{zavm}{m}{n}
\DeclareMathSymbol{\vardiamond}{\mathalpha}{extraup}{87}

\usepackage{framed}

\begin{document}
\title{ThrottleBot - Performance without Insight}
\author{\rm{Michael Alan Chang$^\dag$ Aurojit Panda$^\ddag$ Yuan-Cheng Tsai$^\dag$ Hantao Wang $^\dag$ Scott Shenker$^{\dag\ast}$}\\
\rm{$\dag$ UC Berkeley $\ddag$ NYU $\ast$ ICSI}}

\maketitle
\section{Introduction}
\label{sec:introduction}
Modern web-scale applications are increasingly built by combining sets of microservices, \ie programs running in isolated containers~\cite{container} and interacting with each other through a virtual network. The microservices comprising an application are commonly built by a variety of different developers, and are designed to be reused by different applications. This is in contrast to more traditional monolithic applications, where application functionality was implemented in one (or a few) binaries and were built by a single entity (often a company). Building applications using microservices has several benefits including reduced development time (by encouraging component reuse), increased scalability (allowing individual microservices to be replicated), etc. As a result web companies such as Google~\cite{Burns2016BorgOA}, Uber~\cite{uber}, Netflix~\cite{netflix}, etc. have embraced this architecture and deployed applications comprised of 100s or even 1000s of microservices.

Building applications using microservices has an effect not only on how applications are built, but also on how they are \emph{deployed} and \emph{managed}. When \emph{deploying} such an application, administrators need to configure, provision and launch all of the microservices that comprise an application, and ensure they can communicate with each other. Administrators use \emph{microservice orchestrators}, \eg Kubernetes~\cite{kubernetes}, to simplify application \emph{deployment}. Orchestrators accept as input a description of the application, and then automate the process of launching microservices, configuring connectivity between microservices, providing microservice discovery, etc. Orchestrators have become indispensable tools for building and deploying applications, and are used both by large companies such as Google, and smaller enterprises.

When managing applications, administrators need to change individual microservices to achieve their aims.  \emph{Performance} is a paramount concern when managing deployed applications, and administrators need to frequently \emph{provision} new resources to ensure applications continue to meet performance requirements. In microservice based applications, to effectively provision resources administrators need to determine \emph{what resources} need to be added to \emph{what microservices} to improve application performance.

\emph{Resource provisioning} for microservice based applications remains a manual process, and requires that administrators account for both how additional resources affect an individual microservice's performance, and how its performance impacts application performance as a whole. Efficient resource provisioning therefore requires administrators to acquire a deep understanding of how each microservice is programmed, and about the interactions between microservices within an application -- a challenging prospect when managing large scale systems. As a result, administrators commonly resolve performance problems by uniformly overprovisioning resources across all microservices in an application. Not all microservices can utilize these additional resources, and hence this technique is inefficient. A more sophisticated approach involves identifying resource bottlenecks for each microservice by measuring utilization. As we show later in \S\ref{sec:motivation:utilization}, this is both \emph{ineffective} -- since it cannot account for interactions between microservices in an application -- and \emph{inaccurate} -- since it might mispredict the actual bottleneck. 

In this paper we propose techniques towards automating resource provisioning, \ie for identifying \emph{what} resource on \emph{which} microservice which would result in the greatest impact on application performance. To solve this problem, our technique uses empirical measurements of application performance when hardware resources (\eg CPU cores, disk bandwidth, etc.) allocated to a microservice are \emph{throttled}.

 We have implemented out techniques in \systemname and show results from applying to a variety of systems in \S\ref{sec:evaluation}. \systemname enables efficient push-button resource provisioning for microservice based applications. We envision that in the future our techniques will be incorporated in \emph{orchestrators}, and that \systemname integrated seamlessly with traditional microservice orchestrators can enable true push-button deployment of web-scale applications.
 
 
\section {Background}
\label{sec:background}
We begin by providing some additional background on applications built using microservices, and microservice orchestrators. Microservice based applications are distributed applications built by combining a set of loosely coupled programs, which we refer to as microservices. Each microservice runs in isolation, and is application agnostic, \ie it can be reused across a variety of applications. For example, ELK stack~\cite{elk}, an application used to index and search through system logs, is built by composing three microservices: (a) Elasticsearch~\cite{elasticsearch}, a microservice designed to search through text documents; (b) Kibana~\cite{kibana}, a microservice for visualizing search results; and (c) Logstash~\cite{logstash}, a microservice used to process and transform text logs. Each of these microservices is also used in other contexts with different requirements -- \eg Elasticsearch is also used to implement autocompletion~\cite{elasticsearchuse} and for searching among documents~\cite{elasticsearchrecipe}. We provide more examples of such applications in \S\ref{sec:motivation} and \S\ref{sec:evaluation}.

While they have many benefits, microservice applications are harder to deploy and administer than monolithic applications~\cite{Burns2016BorgOA}. When deploying a monolithic application an administrator must only configure the application (\eg by updating a setting file) and launch the application binary. In contrast when deploying an application built with microservices an administrator needs to (a) configure each individual microservice, (b) configure connectivity between microservices in an application, (c) provide mechanisms through which a microservice can \emph{discover} other microservices, (d) ensure dependent microservices are launched in the correct order, (e) configure mechanisms to respond to partial application failure, etc. Administrators rely on  \emph{microservice orchestrators} to address many of these challenges. Orchestrators -- \eg Kubernetes~\cite{kubernetes}, Terraform~\cite{terraform}, Docker Compose~\cite{dockercompose}, Quilt~\cite{quilt}, etc. -- are responsible for downloading and launching microservices, configuring connectivity, implementing microservice discovery and lifecycle management (\eg detection and responding to failures).

\eat{-- accept as input an application description specifying both the set of microservices in an application, and dependencies between them. They use this information to download and launch appropriate versions of all microservices in an application, and to configure connectivity between these microservices. Orchestrators generally run each microservice within a container~\cite{container, Soltesz2007ContainerbasedOS}, and are responsible for container life cycle management (\eg detecting failures, killing and relaunching containers when required), implementing microservice discovery and configuring connectivity between containers in an application. Increasingly, orchestrators also provide mechanisms that administrators can use to handle application failures~\cite{k8sfailure}.}

\section{Throttlebot Overview} 
\label{sec:motivation}
\systemname addresses the resource provisioning problem which we define as follows: 
Consider an application comprised of microservices $\mu_1,\allowbreak\ \mu_2,\allowbreak\ldots,\allowbreak\mu_n$, and a set of resources $r_0,\allowbreak\ r_1,\allowbreak\ldots,\allowbreak r_m$. Consider a deployment of this application where each microservice $\mu$ is assigned $\alpha_{\mu,r}$ units of resource $r$ initially. Further, assume that an administrator wants to improve some performance metric $P$. A solution to the resource provisioning problem identifies a microservice $\mu$ and resource $r$ that has significant improvements to $P$. We refer to such microservice-resource pairs $(\mu, r)$ as the \emph{Impacted Microservice Resources} (\bnrs). Note that in general (a) an application does not have a unique \bnr -- \ie improvements to several microservice-resource pairs might result in similar improvements to $P$; and (b) in some cases it might be preferable to return an ordered list of \bnrs rather than identify a single \bnr. We also define the \emph{Maximal Impacted Microservice Resource} (\mbnr) to be the unique \bnr that maximizes improvements to $P$.

In current deployments, administrators respond to performance problems by uniformly increasing all resources across all microservices~\cite{Dean2013TheTA}. While overprovisioning resources in this manner is often sufficient to allow application to meet performance requirement, it results in increased costs~\cite{Gupta:2011:DTD:1963559.1963560}. Furthermore, as we show in \S\ref{sec:evaluation}, provisioning additional resources does not always improve application level performance, and might in fact result in worsen performance.

One might consider tools such as Unix \texttt{top}, Linux \texttt{perf}~\cite{Melo2010TheNL}, etc. to measure resource utilization\footnote{Utilization here  refers to measurements such as percentage of memory used, amount of time a CPU is idle, network and disk queue occupancy, etc.} and identify an application's \bnrs. This is insufficient for several reasons: (a) these tools measure resource utilization for a single microservice, and utilizations cannot be compared across microservices (which might run on different machines) and hence cannot be used to identify \bnr; (b) as we show later in \S\ref{sec:motivation:utilization}, utilization is often insufficient for identifying resource bottlenecks;  (c) these tools cannot account for application level dependencies between microservices, and as we show in \S\ref{sec:motivation:utilization:multiservice} these interdependencies can affect application performance; (d) accurately measuring resource utilization using tools like \texttt{perf} requires access to hardware counters which might not be available in all deployments.


In \systemname we use a different approach for determining application level performance and rely on empirical measurements instead of utilization to identify the \bnr. We first show that such a technique is necessary, and that utilization alone is insufficient for identifying an application's \bnr.

\begin{figure*}[th]
\minipage{0.45\textwidth}
    \includegraphics[width=\linewidth]{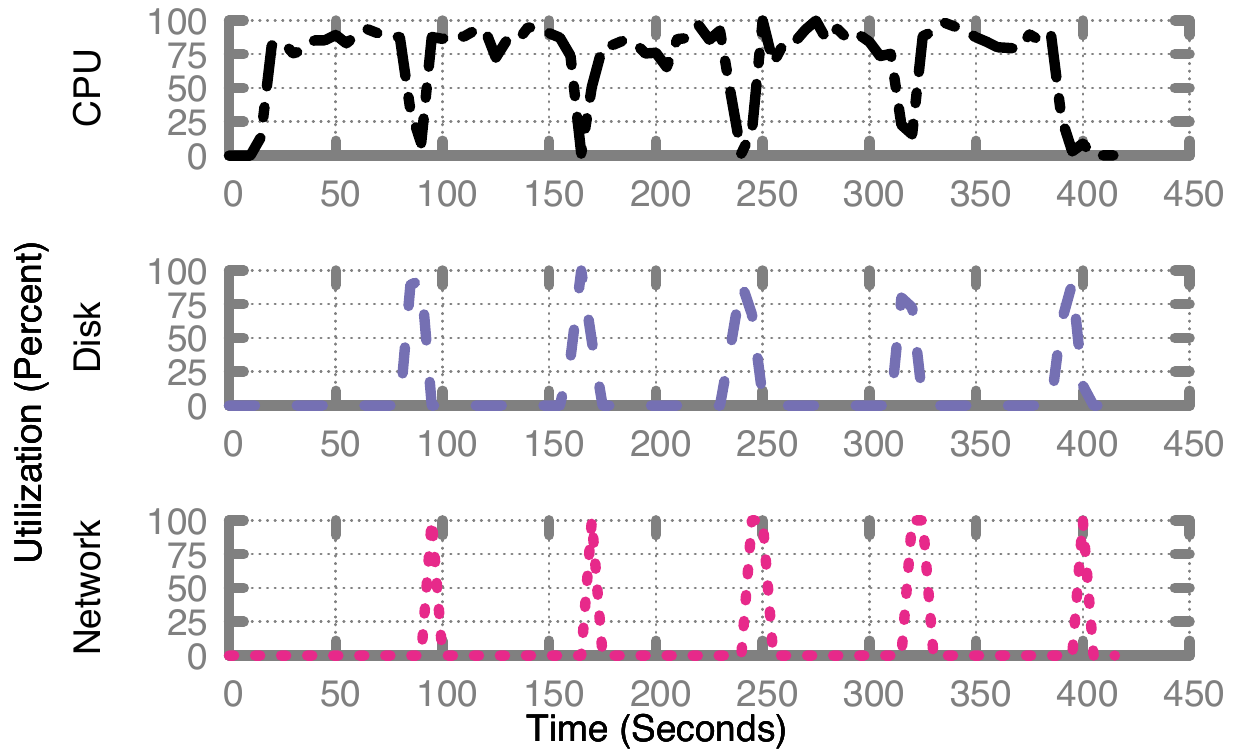}
    \caption{Observed utilization for the single microservice microbenchmark.}
    \vspace{-.1in}
    \label{fig:util-homogenous}
\endminipage\hfill
\minipage{0.45\textwidth}
    \includegraphics[width=\linewidth]{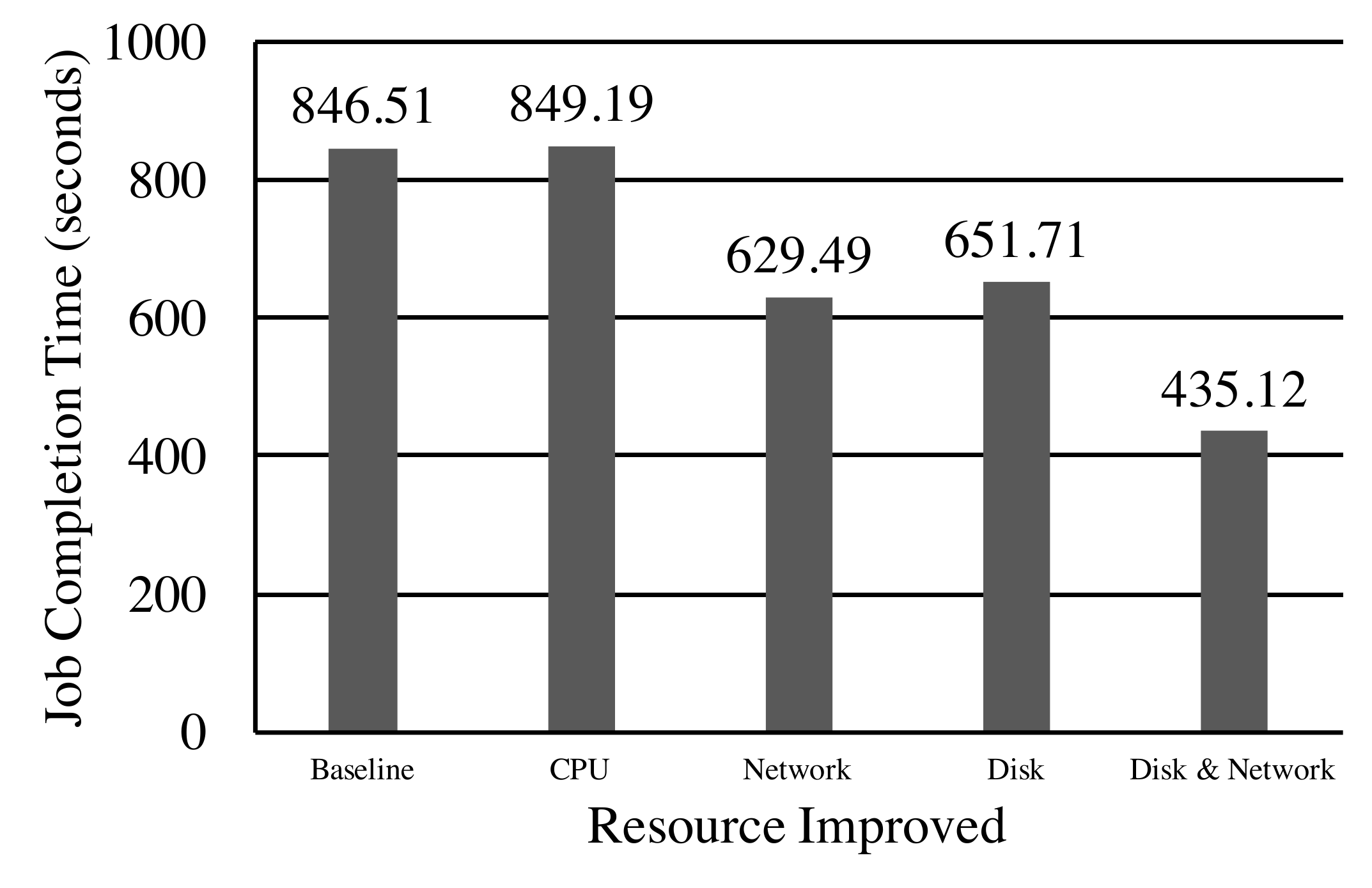}
    \caption{Actual Job Completion time increasing resources allocated to the single microservice microbenchmark.}
    \vspace{-.1in}
    \label{fig:improve-homogenous}
\endminipage\hfill
\end{figure*}

\subsection{Utilization is Insufficient}
\label{sec:motivation:utilization}
As discussed above, utilization cannot always be used to successfully identify an application's \bnr. Below we present two microbenchmarks that illustrate the pitfalls of using utilization to identify \bnr. Through the rest of this paper we focus on three resources: CPU cores, network throughput and disk throughput. Note that our techniques are easily extended to other resources such as memory, however we focus on these three for ease of exposition. We also used \emph{cAdvisor}~\cite{cadvisor} to monitor container resource usage. cAdvisor is widely used to measure utilization in production deployments.

For the following microbenchmarks we started by collecting resource utilization and job-completion time for each microservice during a baseline run. We then measured change in job-completion time when a single microservice is assigned more of a single type of resource -- note that all other microservices and resources are held constant from the baseline. Please see \S\ref{sec:evaluation} for a description of how we add additional resources. We use these measurements to show that the bottleneck resource identified using resource utilization does not correspond to the \bnr in these applications.

\subsubsection{Single Service Microbenchmark}
Our first microbenchmark looks at an application consisting of a single microservice. The microservice computes values, and periodically writes the resulting data to both local disk and remote storage (over the network). To ensure durability, the application also flushes the disk cache after write. This workload closely corresponds to systems like database servers, which flush data to disk for durability~\cite{Pillai2014AllFS}. We show resource utilization for this microservice in Figure~\ref{fig:util-homogenous}.

The baseline utilization graph (Figure~\ref{fig:util-homogenous}) shows high average CPU utilization, which might lead one to conclude that adding \textbf{CPU} cores is most likely to improve job completion time in this application.

However, as can be seen from the Figure~\ref{fig:improve-homogenous}, in reality adding additional CPU cores results in \emph{no improvements} to job completion time. We instead find that the \textbf{disk} and \textbf{network} are the \bnr for this application, and that allocating additional disk and network resources results in a nearly $50\%$ improvement in job completion time (and improving either results in a $26\%$ improvement). This is because the application blocks on disk and network I/O, and adding cores does not help~\cite{Ousterhout2015MakingSO}.


\subsubsection{Multiservice Microbenchmark}
\label{sec:motivation:utilization:multiservice}
\begin{figure*}[ht]
\minipage{0.325\textwidth}
    \includegraphics[width=\linewidth]{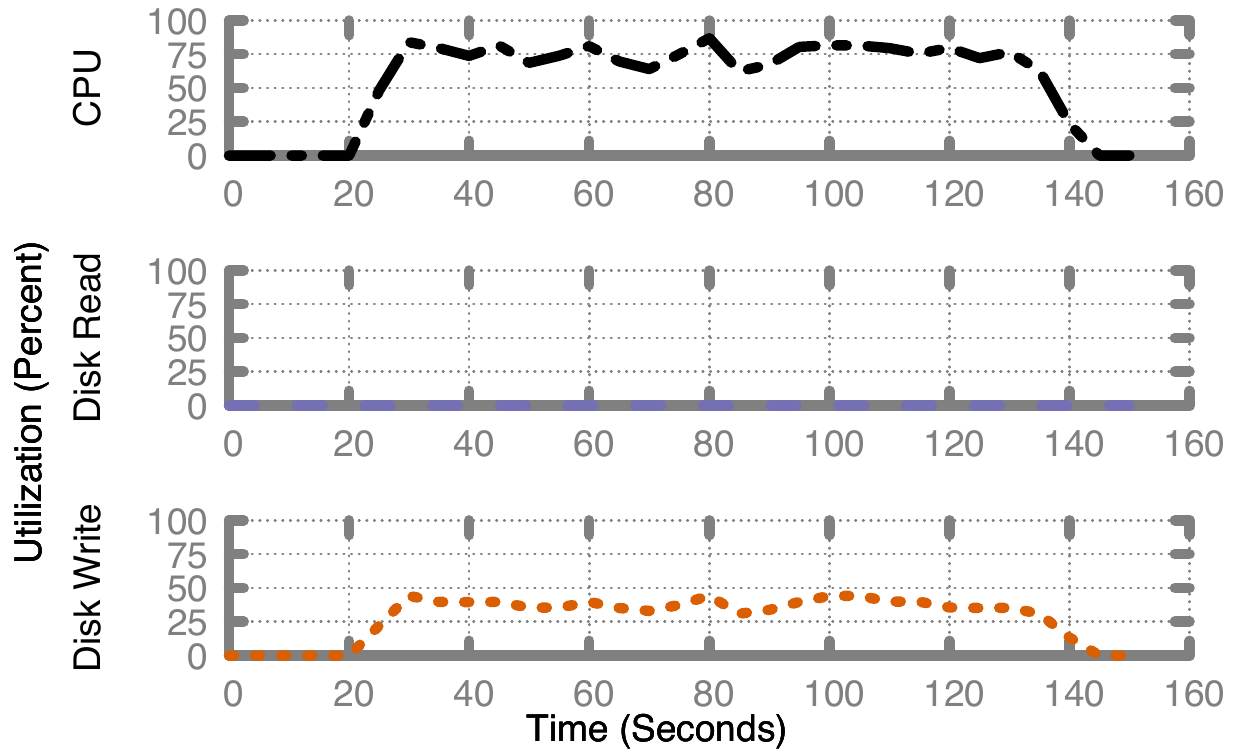}
    \caption{Observed utilization for frontend microservice.}\label{fig:util-flaskapp}
    \vspace{-0.1in}
\endminipage\hfill
\minipage{0.325\textwidth}
    \includegraphics[width=\linewidth]{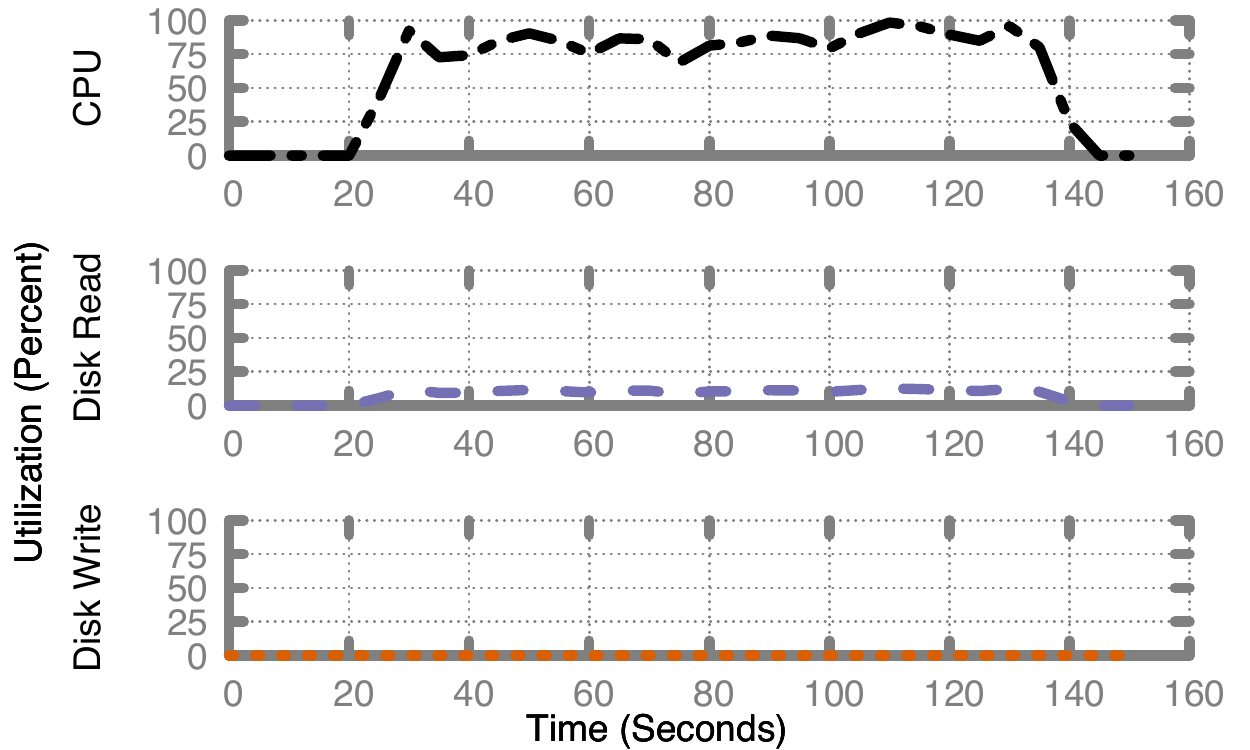}
    \caption{Observed utilization for storage microservice}\label{fig:util-fioapp}
    \vspace{-0.1in}
\endminipage\hfill
\minipage{0.325\textwidth}
    \includegraphics[width=\linewidth]{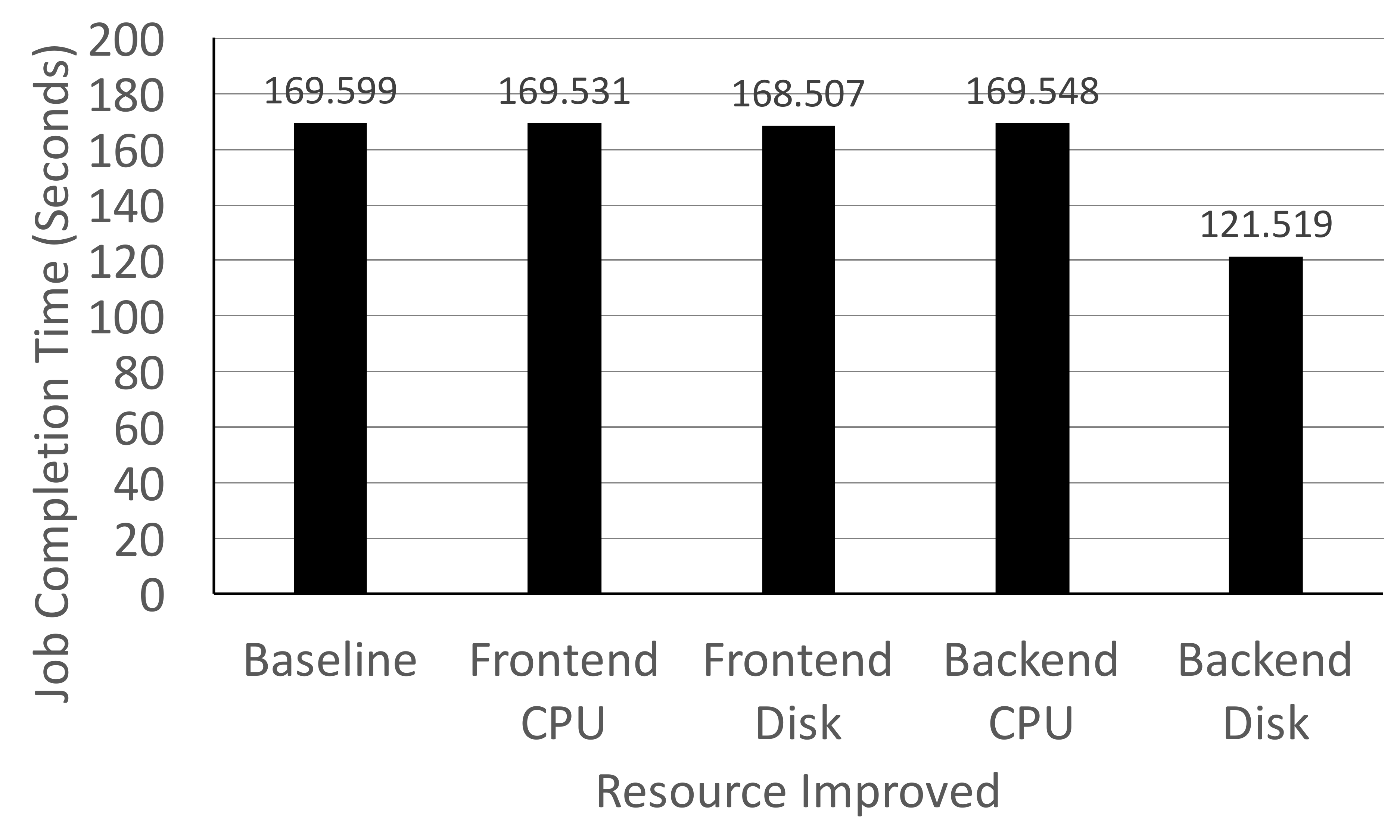}
    \caption{Actual Job Completion Time after adding resources to the multiservice microbenchmark.}\label{fig:hetero-improvement}
    \vspace{-0.1in}
\endminipage
    \vspace{-0.1in}
\end{figure*}

For our second benchmark we use an application comprised of two microservices: (a) a frontend microservice that receives client requests and performs a blocking remote read to the storage microservice; and (b) a storage microservice which on receiving a read request, reads and pre-processes the request file before returning the file's content to the frontend microservice. Figure~\ref{fig:util-flaskapp} and Figure~\ref{fig:util-fioapp} show utilization for the frontend and storage microservices. These figures might lead one to believe that an \bnr could be:
\begin{compactenum}
\item \textbf{CPU} on the \textbf{frontend} microservice;
\item \textbf{CPU} on the \textbf{storage} microservice;
\item \textbf{Disk} on the \textbf{frontend} microservice, which is used for logging.
\end{compactenum}

We measure the impact of adding resources to each microservice in Figure~\ref{fig:hetero-improvement}, and find that none of the above three are an \bnr. Furthermore, we observe no improvements from adding CPU cores to either service, nor from improving disk throughput for the frontend service.

Instead we find that the highest performance impact comes from improving \textbf{disk throughput} on the \textbf{storage microservice}. This is because the frontend microservice cannot respond to a query before the storage microservice is done reading the file. This is an application level dependency that cannot be detected using utilization. This microbenchmark thus demonstrates two problems with using utilization to find an \bnr: (a) utilization cannot be compared across microservices, and hence utilization alone is insufficient to identify the microservice that resources need to be allocated to; and (b) an \bnr might have lower utilization than other resources. Thus we find that existing utilization approaches are insufficient for solving the resource allocation problem.

\section{Throttlebot Approach}
\systemname finds an \bnr empirically by exhaustively stressing application resources, a tried and true method of identifying contended resources in an application. The primary insight that \systemname utilizes is that provisioning more of a resource can be expensive and time-consuming, while throttling a resource is cheap and easily automated. Throttling resources requires neither the purchase of additional hardware nor migrating microservices across physical machines. We discuss the settings for which this premise applies further in \S\ref{sec:discussion}.

The \systemname workflow is simple. At the outset, \systemname accepts as input a configuration file that specifies which microservices, machines, and resources that should be blacklisted for throttling. Based on this configuration, \systemname generates a schedule that will sequentially throttle container resources. Additionally, the administrator provides a pertinent performance metric (\eg job completion time) and produces a representative workload that is used as input to the application; the nature of this workload is further addressed in \S\ref{sec:discussion}. Once the throttling schedule is generated, \systemname will sequentially stress each resource in its stress scheduler and measure how stressing that resource degrades the pertinent performance metric specified by the administrator. Stressed resources that result in the greatest degradation in performance are marked as potential \bnrs, thus providing guidance to administrators on where to add more resources. The administrator can choose to mitigate one of these \bnrs by provisioning more resources to the microservice. Provisioning additional resources might require the administrator to invest in better hardware, or migrate the service to a different machine. The administrators repeats this process until an application meets its performance requirements.

\vspace{-0.1in}
\section{Implementation}
After an orchestrator has deployed a microservice application, \systemname deploys automatically at the press of a button. While \systemname relies on functionality provided by the orchestrator, it does not require special functionality and is agnostic to the orchestrator. As such, we have deployed \systemname with both Kubernetes and Quilt\footnote{\url{quilt.io}}, an open source orchestrator targeted for companies without significant systems expertise. 

Our current implementation makes the following assumptions about the application under test, which we discuss in greater detail in \S\ref{sec:discussion}. 
\begin{asparaenum}
\item Each microservice can fully utilize provisioned resources: We assume each microservice can take advantage of any additional resources it is provided. 
\item Workload: The developer has a representative workload that reflects the actual production workload. 
\item  On-premise: the application is deployed on-premise, where adding more of a resource is extremely costly. 
\end{asparaenum}

\subsection{Resource Throttling}
In this paper we restrict \systemname to throttling a microservice's CPU allocation, disk and network bandwidth.  \systemname can be easily extended to allow stressing of other resource. Below we describe how we stress these resources:\\
\textbf{CPU:} Linux Control groups, or \texttt{cgroups}~\cite{redhat_cgroups} are a kernel feature that enables resource allocation. \systemname currently makes use of two ways to throttle CPU, both using the CPU subsystem. First, \systemname can dictate the proportion of CPU time that the CFS scheduler allocates to a container by setting a container's period and quota. Secondly, \systemname can also add or subtract the number of cores that a container is pinned to through the \texttt{cpuset}~\cite{redhat_cgroups} subsystem.\\
\textbf{Disk Bandwidth:} \systemname throttles CPU through the \texttt{blkio}~\cite{redhat_cgroups} subsystem (again through \texttt{cgroups}), which allows for hard limits on both the read and write from specific block devices. \systemname allows for both joint and individual throttling of read and write. Our experience with the \texttt{blkio} subsystem indicates that using it imposes limited CPU overheads.\\
\textbf{Network Bandwidth:} \systemname stresses the network by limiting link bandwidth using a standard Linux based tool, \texttt{tc}\cite{tc_man}. To do this we first measure the maximum attainable inter-VM network bandwidth, and then impose k\% stress by limiting the container's bandwidth to $(1-k)$\% of this maximum. For example, if a VM is connected using a link with capacity c, \systemname imposes 20\% stress by limiting the network bandwidth to $0.8c$. \texttt{tc} uses hierarchical token bucket (HTB) to implement this rate limit, and scheduling network traffic using HTB imposes some CPU overhead. In our experience this additional overhead did not noticeably affect our results. We could have alternately used an artificial network intensive job to impose stress, however this was set aside because of the additional CPU overhead imposed.\\

\vspace{-0.1in}
\section{Evaluation}
\label{sec:evaluation}
\systemname was deployed on a variety of microbenchmarks and realistic applications, the results of which are described below. Similar to the microbenchmarks described earlier (\S\ref{sec:motivation:utilization}) we applied \systemname to applications deployed using Quilt on Amazon's EC2. The various \bnrs proposed by \systemname are validated by comparing those results against a manual improvement of resources. Note that \systemname proposes several \bnrs but does not offer suggestions on how much to increase those resources by. If the hardware resources were fully provisioned to the services in the baseline experiment, we upgraded the resource by deploying the full set of containers on an upgraded machine (typically the next level up in the instance family), while throttling back non-\bnr resources to the same allocation on the machine where the baseline experiments were conducted. If the hardware resources were not fully provisioned to the services in the baseline experiment, the resource upgrade consisted of increasing the resource allocation to use up remaining unused provisions on that machine. The implications of \textit{how much} to improve an \bnr is further discussed in \S\ref{sec:discussion}.

\subsection{Microbenchmark}
\begin{figure}[t]
    \centering
    \includegraphics[width=0.42\textwidth]{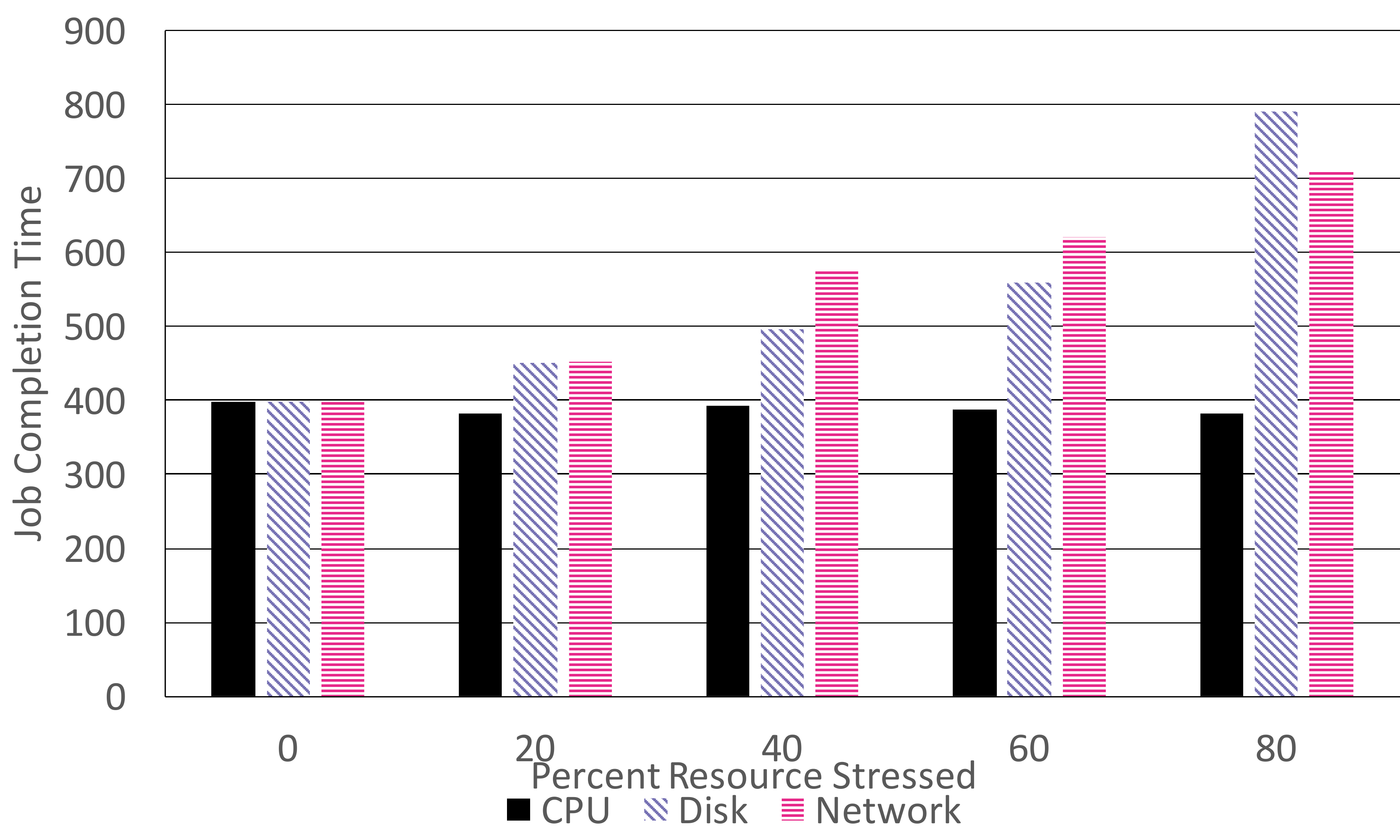}
    \caption{Effect of stressing the Single Service Microbenchmark using \systemname.}\label{fig:tbot-homogenous}
\end{figure}

We revisit the single service, homogeneous microbenchmark from \S\ref{sec:motivation} and demonstrate that \systemname finds an \bnr. In \S\ref{sec:motivation}, we identified the \bnr through an exhaustive and tedious manual improvement of the resources. Running \systemname on the microbenchmark yields the results shown in Figure \ref{fig:tbot-homogenous} automatically, without any administrator intervention after \systemname is initiated. Relative to the baseline (indicated by 0\% stressed), stressing disk and network resulted in a performance degradation while CPU remained constant. \systemname offers two clear and accurate signals that suggest two possible \bnrs. In this case, \systemname does not suggest a clear \mbnr, as network stress degrades performance more at 40\% stress, while disk stress degrades performance more at 80\% stress. While this microbenchmark is a special case where improving both resources result in similar improvements, the application benefits equally from provisioning more network and disk. We discuss the interpretability of \systemname results later in \S\ref{sec:discussion}. Ultimately, \systemname accurately proposes two \bnrs and offers a clear and accurate signal that adding more network and disk would serve to improve the performance of the application the most. 

\subsection{Realistic Applications}

\subsubsection{Spark-Streaming Pipeline}
\label{sec:evaluation:stream}
It is increasingly common to deploy big data platforms in a streaming pipeline, which ingest, process, and store real time data. One such widely deployed streaming pipeline consists of Kafka (a publish-subscribe messaging microservice), Spark Streaming (a platform for processing streams of data), and Redis (an in-memory data store) \cite{sparkstreaming_pipeline}. We deployed this streaming pipeline with the Kafka microservice, a Spark cluster consisting of 8 workers nodes, and Redis. We ran the well-known Yahoo Streaming Benchmark, which is an advertising application specifically designed in industry to evaluate the performance of streaming platforms on realistic operations \cite{yahoo_benchmark}. We subsequently used \systemname on this deployment, measuring the impact of stresses on the event latency, which uses window update times to measure the latency required of going from the event generator to Kafka before being written to Redis by the Spark Streaming job. With a sending workload of 50,000 requests per second, the change in latency as a result of using \systemname are shown in Table \ref{table:spark-streaming}. Note that while Spark is considered to be a single microservice, \systemname automatically considers the Spark master and the Spark worker to be different services in the application because of the vastly different role they play even within the same microservice.
    
As shown in bold, \systemname automatically detects two strong signals, identifying the $(\text{kafka},\ \text{network})$ tuple as the \mbnr and the $(\text{spark-master},\ \text{Disk})$ tuple as another strong candidate as an \bnr, due to their stresses causing significant increases in latency. Upon provisioning more network to Kafka, the \mbnr, window latency improved from the baseline of 31.95 seconds to 16.58 seconds, an improvement of nearly 2x. The other \bnr resulted in a performance improvement from 31.95 seconds to 25.9 seconds, an improvement of 20\%. Indeed, after manually improving resources on every other $(resource, service)$ tuple, improving the network on Kafka proved to be the most impactful. Thus \systemname was able to successfully identify the \mbnr and another \bnr, and offer suggestions on what other resources might offer meaningful improvements.\\

While \systemname was effective on this level, our manual process of improving a single resource at a time revealed that \systemname failed to identify certain resources that, when improved, would have resulted in significant improvements in performance. While these resources were not an \bnr, \systemname should still have signaled that resource as a \bnr. For instance, throttling the number of CPU cores on Kafka had no impact on the window latency; in fact, Table \ref{table:spark-streaming} shows the window latency slightly decreased when the number of cores is reduced. Despite this, our exhaustive search suggested that improving this resource would result in the a 40\% improvement in performance. We discuss this gap in \systemname's report in \S\ref{sec:discussion}.

\eat{The evaluation of this streaming pipeline revealed that there exist many cases where applying a small stress to a resource can greatly \textit{improve} overall end-to-end event latency. For example, stressing a resource using \systemname sometimes reduces the window update latency in the Spark Streaming application. For example. provisioning less network bandwidth to a container results in a performance improvement, from 31.94 seconds to 18.56 seconds, almost a 2x improvement. Our other experiments have indicated that this is not an uncommon scenario in Spark jobs; injecting network latency or reduced network throughput to a Spark cluster often reduces premature sending of tasks which interrupt still ongoing remote processes. Similar phenomenon have previously been observed in other concurrent applications~\cite{Scherer2005AdvancedCM, Gidra2013ASO}.}

\begin{table}[t]
\centering
\resizebox{\columnwidth}{!}{%
    \begin{tabular}{|l|r|r|r|r|r|r|}
        \hline
        Microservice & CPU Cores & Disk & Network\\
        \hline
        \hline
        Spark-master & -3242 &  \textbf{+25578.4} & +1922.4 \\
        \hline
        Spark-worker & +546.3 & +2433.9 & +9547.7  \\
        \hline
        Kafka & -2869.3 & +4058.2 & \textbf{+98696.2} \\
        \hline
        Redis & +488.5 & -1552.9 & +7481 \\
        \hline
    \end{tabular}
    }\caption{Change in latency (in ms) upon throttling in Streaming Pipeline. Bolded figures indicate the application's \bnr.}
    \vspace{-0.1in}
\label{table:spark-streaming}
\end{table}    
    
\subsubsection{Mean Stack}
Many simple web applications deployed today are deployed with the MEAN software bundle \cite{mean_stack}. Applications using this bundle typically require deploying three microservices: load balancer, web server, database. We deployed the MEAN stack with HAProxy Load Balancer, Nginx Web Server, and MongoDB database. \systemname identifies the \bnr as the CPU resource on the Nginx container, regardless of a number of metrics that an administrator might be concerned with: p99 request latency, request throughput, median latency, and HTTP failure rate. Provisioning additional cores to the Web server resulted in a 35\% improvement in p99 latency, while provisioning more resources to other services resulted in a negligible improvement in performance

\section{Discussion}
\label{sec:discussion}
Next, we discuss the appropriateness of \systemname's assumptions, some limitation, and avenues for future work.

\noindent\textbf{Relating performance degradation and improvement:} Currently \systemname assumes a correlation between performance degradation when a resource is throttled, and performance improvements from increasing that resource. In our experiments we found that this correlation held across a variety of applications, but we plan to add feedback policies to handle applications where this is not the case.

\noindent\textbf{Can \systemname always identify the \bnr?} While in our experiments \systemname always correctly identified an \bnr, we have found that the efficacy of our technique depends both on \emph{how} resources are throttled and by \emph{how much}. For example, for the streaming benchmark (\S\ref{sec:evaluation:stream}) we found that (a) provisioning additional cores resulted in performance improvements even when throttling resulted in no degradation (b) identifying an \bnr in this case is easier when we reduce CPU quotas as opposed to cores. In future work we plan to both integrate several methods for stressing each resource, and work on techniques to identify the amount of stress that should be applied.


\noindent\textbf{Paradoxical Results}: During our experiments, \systemname uncovered cases where stressing resources produced paradoxical results where performance improved. For example, we found that when running \systemname on an application built using the Spark distributed matrix library \cite{Sparks2017KeystoneMLOP}, injecting a small amount of network latency resulted in job completion time improvement by approximately 12 percent. Injecting network latency or reducing network throughput in a Spark cluster reduces the rate at which the driver sends processes, and hence minimizes interruption to an ongoing remote processes. Similar phenomenon have previously been observed in other concurrent applications~\cite{Scherer2005AdvancedCM, Gidra2013ASO}. These paradoxical results can help the administrator more effectively determine more placements and co-location policies of various services.

\noindent\textbf{Microservices make use of all provisioned resources:} We assume that microservices automatically take advantage of all provisioned resources. However, a microservice's configuration might change based on resource availability. For these cases we rely on a user supplied script which can appropriately update the configuration.

\noindent\textbf{Workload Generation}: An application's \bnrs depends on its workload, and \systemname is sensitive to workload changes. Identifying a representative workload is non-trivial, and has been the subject of recent work (\eg~\cite{Veeraraghavan2016KrakenLL}). Workloads for offline application (\eg analytics applications) tend to be fairly stable and can be derived from past jobs. Online services have more variable demands due to variances from the diurnal pattern, etc. In the future we plan to investigate techniques for safely deploying \systemname in production environments, thus eliminating the need to generate a workload.

\noindent\textbf{Deployment Setting}: \systemname injects stresses into a system on the premise that adding additional resources to a system is relatively expensive -- both in time and resources. As such, \systemname is particularly effective for the on-premise deployment, where system administrators need to make difficult decisions about what hardware to invest in. While \systemname can be deployed effectively in the cloud environment, actually adding resources to an application is a much smaller investment. However, an individual machine might host thousands of containers~\cite{density}, and migrating all containers to a new machine is prohibitive, especially in production. To safely apply \systemname in production environments, we plan on adding the ability to measure the impact of adding resources to a container if they are available; otherwise, it will apply the throttling approach described in this paper.

\noindent\textbf{Pruning Search Space:} Our current implementation performs an exhaustive search by stressing all resources allocated to all microservices. This might be impractical for large applications, \eg applications with 100s or 1000s of microservices. We are currently investigating mechanisms to prune the search space so that we can apply \systemname to large applications.

\noindent\textbf{Choice of resources:} In this paper we have focused on three resources: CPU cores, disk and network throughput. We focused on these resources for ease of experimentation and exposition. However our techniques are general and can easily be extended to resources such as memory, GPU time, etc.

\noindent\textbf{Provisioning additional replicas vs adding resources:} \systemname focuses on finding \bnrs and allowing administrators to add resources to a running microservice. An alternative strategy for scaling would be to launch new replicas of a microservice. We did not investigate this strategy since replicating microservices requires application support, and might not be possible in general.

\section{Related Work}
Several systems have proposed mechanisms for provisioning resources in compute clusters (\eg choosing optimal EC2 instance type), but they assume a particular software framework \cite{Venkataraman:2016:EEP:2930611.2930635} or that resources are provisioned equally between all services in the cluster \cite{Yadwadkar:2017, Alipourfard2017CherryPickAU}. Proposed systems allow operators to infer performance behaviors of various systems, but they largely require significant modifications to the VMM or to the application under test \cite{Gupta:2011:DTD:1963559.1963560, Gupta:2005:IBT:1095810.1118605} or only explore performance improvements along a single resource dimension (e.g., network) \cite{Gupta:2005:IBT:1095810.1118605,Pan2005SHRiNKAM,Bucy2008TheDS}.  
Other works have suggested improving application performance through profiling programs and optimizing code \cite{Curtsinger:2015:COF:2815400.2815409, Miller1995ThePP, Burtscher2010PerfExpertAE}; these provide a different set of knobs from \bnr, and thus can be jointly used with \systemname to improve application performance. Past proposals on resource scheduling \cite{Ghodsi:2011:DRF:1972457.1972490, Mao:2016:RMD:3005745.3005750, Grandl2014MultiresourcePF} assume that the administrator provides resource requirements as input; thus we view this work as complimentary to \systemname.

\section{Conclusion}

Microservice orchestrators have greatly simplified the process of deploying large scale web-applications, thus enabling their deployment by administrators with varying levels of expertise. However, once deployed, provisioning these applications to meet performance goals remains challenging. In this paper we proposed \systemname, which automates provisioning, enabling \emph{push-button} deployment and management of web-scale applications.

\newpage
\bibliographystyle{abbrv}
\begin{small}
\bibliography{tbot}
\end{small}
\end{document}